\documentstyle[12pt]{article}
\begin{document}

\begin{flushright}
PITHA 94/19\\
hep-ph/9404259\\
March, 1994
\end{flushright}
\vspace*{2cm}
\LARGE\centerline{$W^+-H^+$ Interference and Partial Width}
\centerline{ Asymmetry in Top and Antitop Decays}
\vspace*{1cm}
\large\centerline{ Torsten Arens and L.M. Sehgal\footnote{E-mail addresses: 
\{arens,sehgal\}@acphyz.hep.rwth-aachen.de}}
\centerline{III. Physikalisches Institut (A), RWTH Aachen,}
\centerline{D-52074 Aachen, Germany}
\normalsize

\begin{abstract}
We re-examine the question of a possible difference in the partial 
decay widths of $t$ and $\overline t$, induced by an intermediate 
scalar boson $H^+$ with $CP$-violating couplings. The interference of 
$W^+$ and $H^+$ exchanges is analysed by constructing the $2\times 2$ 
propagator matrix of the $W^+-H^+$ system, and determining its 
absorptive part in terms of fermion loops. Results are obtained for 
the partial rate difference in the channels $t\to bl^+\nu_l $ and 
$t\to bc\overline s$, which fulfil explicitly the constraints of $CPT$ 
invariance. These results are contrasted with those in previous work.
\end{abstract}

\newpage
\section{Introduction}
Recent literature \cite{atwood,liu} has been witness to 
an interesting debate on the question of a possible $CP$-violating 
difference in the partial widths of $t$ and $\overline t$ decays, into 
conjugate channels such as $t\to b\tau^+\nu_{\tau}$ and $\overline t\to
\overline b\tau^-\overline{\nu}_{\tau}$. This discussion has  
taken place in the context of a model in which the decays of the top 
quark are mediated, not only by $W^{\pm}$ bosons, but also by charged 
Higgs bosons $H^{\pm}$ with $CP$-violating couplings \cite{albright}. 
Two specific questions that have arisen in this regard are (i) the 
correct form of the propagator for an unstable $W$ boson \cite{atwood,
liu,nowakowski,lopez}, and (ii) the implications of $CPT$ invariance and
 unitarity for partial rate asymmetries generated by absorptive parts 
of decay amplitudes \cite{wolfenstein}.\newline
\indent In this paper, we present an analysis that, we believe, is more 
complete than that in Refs. \cite{atwood,liu}. Central to our analysis 
is the derivation of the propagator matrix of the coupled $W^+-H^+$ 
system, taking account of vacuum polarization effects induced by 
fermion loops. The propagator matrix includes off-diagonal transitions 
between $W^+$ and $H^+$, which turn out to be essential for obtaining 
a partial rate asymmetry that respects the constraints of $CPT$ 
invariance. \newline
\indent The model we use is defined by the Lagrangian \cite{albright}
\begin{eqnarray}
\cal L&=&{\cal L}_W+{\cal L}_H,\\
{\cal L}_W&=&-\frac{g}{2\sqrt 2}\biggl\{\sum_{l=e,\mu ,\tau}\Bigl [\overline 
\nu_l\gamma^{\mu}(1-\gamma_5)lW_{\mu}^++\overline l\gamma^{\mu}(1-
\gamma_5)\nu_lW_{\mu}^-\Bigr ]\nonumber\\
&&+\overline u\gamma^{\mu}(1-\gamma_5)dW_{\mu}^++\overline d
\gamma^{\mu}(1-\gamma_5)uW_{\mu}^-\nonumber\\
&&+(u,d)\to (c,s)+(u,d)\to (t,b)\biggr\},\\
{\cal  L}_H&=&\frac{g}{2\sqrt 2}\biggl\{-\sum_{l=e,\mu ,\tau}\Bigl [H^+Z
\frac{m_l}{m_W}\overline {\nu}_l(1+\gamma_5)l+H^-Z^*\frac{m_l}{m_W}
\overline l(1-\gamma_5)\nu_l\Bigr ]\nonumber\\
&&+H^+\overline u\Bigl [Y\frac{m_u}{m_W}(1-\gamma_5)+X\frac{m_d}{m_W}
(1+\gamma_5)\Bigr ]d\nonumber\\
&&+H^-\overline d\Bigl [X^*\frac{m_d}{m_W}(1-\gamma_5)+Y^*\frac{m_u}
{m_W}(1+\gamma_5)\Bigr ]u\nonumber\\
&&+(u,d)\to (c,s)+(u,d)\to (t,b)\biggr\},
\end{eqnarray}
where we neglect quark-mixing. The parameters $X,Y,Z$ appearing in ${\cal L}_H$ 
are permitted to be complex relative  to one another, so that this term 
is $CP$-violating. The interaction ${\cal L}_H$ may be imagined to arise as a 
special case of the Weinberg model with three Higgs doublets \cite{
weinberg}, in which the remaining charged scalars are sufficiently 
heavy to be disregarded.

\section{The $W^+-H^+$ Propagator}
We are concerned with the propagator (in unitary gauge) of the coupled 
$W^+-H^+$ system, which we describe by a $2\times 2$ matrix
\begin{eqnarray}
D=\left (\begin{array}{cc}D_W^{\mu\nu}&D_{W^+H^+}^{\mu}\\
D_{H^+W^+}^{\nu}&D_H
\end{array}\right ).
\end{eqnarray}
The inverse of this matrix is defined by 
\begin{eqnarray}
D^{-1}D=\left (\begin{array}{cc}g&0\\
0&1\end{array}\right ),
\end{eqnarray}
where $g$ is the metric tensor with elements $g^{\mu\nu}=diag(1,-1,-1,-1)$.
The inverse matrix $D^{-1}$ has the general form
\begin{eqnarray}
D^{-1}=-i\left (\begin{array}{cc}
(m_W^2-q^2+F_1)g^{\mu\nu}+q^{\mu}q^{\nu}(1+F_2)&q^{\mu}F_3\\
q^{\nu}F_4&q^2-m_H^2+F_5
\end{array}\right ),
\end{eqnarray}
where the functions $F_i(q^2)$, $i=1,\cdots ,5$ are given by the 
one-particle-irreducible self-energies
\begin{eqnarray}
\Sigma_W^{\mu\nu}(q^2)&=&i\Bigl [g^{\mu\nu}F_1(q^2)+q^{\mu}q^{\nu}F_2(q^2)
\Bigr ],\nonumber\\
\Sigma_{W^+H^+}^{\mu}(q^2)&=&iq^{\mu}F_3(q^2),\nonumber\\
\Sigma_{H^+W^+}^{\mu}(q^2)&=&iq^{\mu}F_4(q^2),\nonumber\\
\Sigma_H(q^2)&=&iF_5(q^2).
\end{eqnarray}
Inversion of the matrix (6) yields the elements of the propagator matrix (4):
\begin{eqnarray}
D_W^{\mu\nu}&=&i\frac{-g^{\mu\nu}+q^{\mu}q^{\nu}\displaystyle\frac{(1
+F_2)(q^2-m_H^2+F_5)-F_3F_4}{(m_W^2+F_1+q^2F_2)(q^2-m_H^2+F_5)-q^2F_3F_4}}
{q^2-m_W^2-F_1},\nonumber\\
D_{W^+H^+}^{\mu}&=&\frac{iq^{\mu}F_3}{q^2F_3F_4-(q^2-m_H^2+F_5)(m_W^2
+F_1+q^2F_2)},\nonumber\\
D_{H^+W^+}^{\mu}&=&\frac{iq^{\mu}F_4}{q^2F_3F_4-(q^2-m_H^2+F_5)(m_W^2
+F_1+q^2F_2)},\nonumber\\
D_H&=&\frac{i}{q^2-m_H^2+F_5-\displaystyle\frac{q^2F_3F_4}{m_W^2+F_1+
q^2F_2}}.
\end{eqnarray}
The corresponding propagator matrix for $W^--H^-$ is obtained by the 
replacement $q^{\mu}\to -q^{\mu}$, $F_3(q^2)\leftrightarrow F_4(q^2)$. The 
above derivation is analogous to the description of the $\gamma -Z$ system 
\cite{hollik}. A graphical representation of Eqs. (7) and (8) is given in 
Figs. 1-3. \\
\indent The function $D_W^{\mu\nu}$, representing the $WW$ element of the 
propagator matrix, can be decomposed into transverse and longitudinal 
pieces:
\begin{eqnarray}
D_W^{\mu\nu}=i(-g^{\mu\nu}+\frac{q^{\mu}q^{\nu}}{q^2})G_T+i\frac{q^{\mu}q^{\nu}}
{q^2}G_L
\end{eqnarray}
with
\begin{eqnarray}
G_T&=&\frac{1}{q^2-m_W^2-F_1},\nonumber\\
G_L&=&\frac{1}{m_W^2+F_1+q^2F_2-\displaystyle\frac{q^2F_3F_4}{q^2-m_H^2+F_5}}.
\end{eqnarray}
It will turn out that only the longitudinal part $G_L$ contributes to the 
partial width asymmetry. If the term proportional to $F_3F_4$ is dropped, 
the function $G_L$ coincides with that in Refs. \cite{atwood,liu}. We work 
initially with 
the full expression in Eq. (8), in order to obtain results that are also 
valid for $q^2\simeq m_H^2$, a region that is physically accessible if 
$m_H<m_t-m_b$.

\section{Difference of Partial Widths}
\subsection{Asymmetry in Lepton Channels}
The amplitude of the decay $t\to bl^+\nu_l$, including vacuum 
polarization effects in the $W-H$ propagator, is given by the sum of the 
four diagrams shown in Fig. 4, and has the form
\begin{eqnarray}
M_l=\frac{ig^2}{8}&\Bigl\{&A_l\overline u_b\gamma^{\mu}(1-\gamma_5)u_t
\overline u_{\nu}\gamma_{\mu}(1-\gamma_5)v_l\nonumber\\
&+&B_l\overline u_b(1+\gamma_5)u_t\overline u_{\nu}(1+\gamma_5)v_l
+D_l\overline u_b(1-\gamma_5)u_t\overline u_{\nu}(1+\gamma_5)v_l\Bigr \}
\end{eqnarray}
with
\begin{eqnarray}
A_l&=&G_T,\nonumber\\
B_l&=&\frac{m_tm_l}{m_W^2}\Bigl \{\frac{m_W^2}{q^2}(G_T+G_L)
+Y^*ZG_5+m_WN(Y^*F_4+ZF_3)\Bigr\},\nonumber\\
D_l&=&\frac{m_bm_l}{m_W^2}\Bigl \{-\frac{m_W^2}{q^2}(G_T+G_L)
+X^*ZG_5+m_WN(X^*F_4-ZF_3)\Bigr\},
\end{eqnarray}
where $N\equiv [(m_W^2+F_1+q^2F_2)(q^2-m_H^2+F_5)-q^2F_3F_4]^{-1}$ and $
G_5\equiv N(m_W^2+F_1+q^2F_2)$.
The corresponding decay amplitude for $\overline t\to\overline bl^-\overline 
{\nu}_l$ is
\begin{eqnarray}
\overline M_l=\frac{ig^2}{8}&\Bigl\{&\overline A_l\overline v_t\gamma^{\mu}
(1-\gamma_5)v_b\overline u_l\gamma_{\mu}(1-\gamma_5)v_{\nu}\nonumber\\
&+&\overline B_l\overline v_t(1-\gamma_5)v_b\overline u_l(1-\gamma_5)
v_{\nu}+\overline D_l\overline v_t(1+\gamma_5)v_b\overline u_l
(1-\gamma_5)v_{\nu}\Bigr\}
\end{eqnarray}
with
\begin{eqnarray}
\overline A_l&=&G_T,\nonumber\\
\overline B_l&=&\frac{m_tm_l}{m_W^2}\Bigl \{\frac{m_W^2}{q^2}(G_T+G_L)
+YZ^*G_5+m_WN(YF_3+Z^*F_4)\Bigr\},\nonumber\\
\overline D_l&=&\frac{m_bm_l}{m_W^2}\Bigl \{-\frac{m_W^2}{q^2}(G_T+G_L)
+XZ^*G_5+m_WN(XF_3-Z^*F_4)\Bigr\}.
\end{eqnarray}
The matrix elements $M_l$ and $\overline M_l$ yield the following 
asymmetry between the partial widths:
\begin{eqnarray}
\Delta_{l\nu_l}&\equiv &\Gamma(\overline t\to\overline bl^-\overline 
{\nu}_l)-\Gamma(t\to bl^+\nu_l)\nonumber\\
&=&\frac{1}{2m_t}\int\frac{d^3p_b}{(2\pi )^32E_b}\frac{d^3p_l}{(2\pi )^3
2E_l}\frac{d^3p_{\nu}}{(2\pi )^32E_{\nu}}(2\pi )^4\delta^{(4)}(p_t-p_b
-p_l-p_{\nu})\nonumber\\
&&\qquad\cdot\Bigl\{\overline{|\overline M_l|^2}-\overline{|M_l|^2}\Bigr\}\nonumber\\
&=&\frac{g^4}{2^{11}\pi^3m_t^3}\int\limits_{m_l^2}^{(m_t-m_b)^2}\frac{dq^2}{q^4}
\lambda (q^2,m_t^2,m_b^2)(q^2-m_l^2)^2\Bigl\{\nonumber\\
&&(|\overline B_l|^2-|B_l|^2+|\overline D_l|^2-|D_l|^2)
q^2(m_t^2+m_b^2-q^2)\nonumber\\
&&+4Re(\overline B_l^*\overline D_l-B_l^*D_l)m_tm_bq^2\nonumber\\
&&-2ReA^*(\overline B_l-B_l)m_tm_l(m_t^2-m_b^2-q^2)\nonumber\\
&&-2ReA^*(\overline D_l-D_l)m_bm_l(m_t^2-m_b^2+q^2)\Bigr \},
\end{eqnarray}
where $\lambda (a,b,c)=(a^2+b^2+c^2-2ab-2ac-2bc)^{1/2}$. Substituting the 
expressions for $A_l,B_l,D_l,\overline A_l,\overline B_l,\overline D_l$ in 
the above integrand, we find (as anticipated) that terms proportional to the 
transverse propagator $G_T$ cancel completely. 
Expressed in terms of the functions $F_i(q^2)$, the asymmetry 
involves only the quantities $Im(F_1+q^2F_2)(q^2)$, $(F_3-F_4^*)(q^2)$ and 
$ImF_5(q^2)$. 
Representing the self-energies by fermion loops, these terms are
\begin{eqnarray}
Im(F_1+q^2F_2)(q^2)\!\!\!&=&\!\!\!\frac{g^2}{16\pi}\Bigl\{\frac{N_c\lambda(q^2,
m_u^2,m_d^2)}{2q^4}\Theta [q^2-(m_u+m_d)^2]\nonumber\\
&&\cdot (-m_u^4-m_d^4+q^2m_u^2+q^2m_d^2+2m_u^2m_d^2)\nonumber\\
&&+(u,d)\to (c,s)+\sum_{l=e,\mu ,\tau}\frac{(q^2-m_l^2)^2}{2q^4}m_l^2
\Theta (q^2-m_l^2)\Bigr\},\nonumber\\
(F_3-F_4^*)(q^2)\!\!\!&=&\!\!\!\frac{g^2}{16\pi}\Bigl\{\frac{iN_c\lambda (q^2,
m_u^2,m_d^2)}{q^4}\Theta [q^2-(m_u+m_d)^2]\nonumber\\
&&\bigl [X^*\frac{m_d^2}{m_W}(q^2+m_u^2-m_d^2)
-Y^*\frac{m_u^2}{m_W}(q^2-m_u^2+m_d^2)\bigr ]\nonumber\\
&&+(u,d)\to (c,s)+\sum_{l=e,\mu ,\tau}\frac{i(q^2-m_l^2)^2}{q^4}(-Z^*)
\frac{m_l^2}{m_W}\Theta (q^2-m_l^2)\Bigr\},\nonumber\\
ImF_5(q^2)\!\!\!&=&\!\!\!\frac{g^2}{16\pi}\Bigl \{\frac{N_c\lambda (q^2,m_u^2,m_d^2)}
{2q^2}\bigl [(|Y|^2\frac{m_u^2}{m_W^2}+|X|^2\frac{m_d^2}{m_W^2})
(q^2-m_u^2-m_d^2)\nonumber\\
&&-\frac{m_u^2m_d^2}{m_W^2}4ReXY^*\bigr ]\Theta
[q^2-(m_u+m_d)^2]+(u,d)\to (c,s)\nonumber\\
&&+\sum_{l=e, \mu ,\tau}\frac{(q^2-m_l^2)^2}{2q^2}|Z|^2
\frac{m_l^2}{m_W^2}\Theta (q^2-m_l^2)\Bigr \}.
\end{eqnarray}
One finds that the contribution of the lepton loops (the pieces $
\displaystyle\sum_{l=e,\mu ,\tau}$) to the asymmetry vanishes identically, 
leaving as the final result 
\begin{eqnarray}
\Delta_{l\nu_l}&=&\frac{g^6N_cm^2_l}{2^{14}\pi^4m_t^3m_W^6}
\int\limits^{(m_t-m_b)^2}_{Max[(m_c+m_s)^2,m_l^2]}\frac{dq^2}{q^6}\frac{\lambda (q^2,
m_t^2,m_b^2)\lambda (q^2,m_c^2,m_s^2)(q^2-m_l^2)^2}{(q^2-m_H^2)^2
+m_H^2\Gamma_H^2}\nonumber\\
&\Biggl\{ &\Bigl [m_t^2m_s^2(m_t^2-m_b^2-q^2)(q^2+m_c^2-m_s^2)\nonumber\\
&&\qquad\qquad\qquad -m_b^2m_c^2(m_b^2-m_t^2-q^2)(q^2+m_s^2-m_c^2)\Bigr ]\nonumber\\
&&\cdot \Bigl [(1-\frac{m_H^2}{q^2})Im(XY^*-XZ^*-YZ^*)\nonumber\\
&&\qquad\qquad\qquad +|Y|^2ImXZ^*-|Z|^2ImXY^*+|X|^2ImYZ^*\Bigr ]\nonumber\\
&&+2ImYZ^*|X+Y|^2m_t^2m_c^2\Bigl [q^2(m_s^2-m_b^2)+m_b^2m_c^2-m_t^2
m_s^2\Bigr ]\nonumber\\
&&+2ImXZ^*|X+Y|^2m_b^2m_s^2\Bigl [q^2(m_t^2-m_c^2)+m_b^2m_c^2-m_t^2
m_s^2\Bigr ]\Biggr\}\nonumber\\
&+&(c,s)\to (u,d)\nonumber\\
&\equiv &\Delta_{l\nu_l}(c,s)+\Delta_{l\nu_l}(u,d).
\end{eqnarray}
We have introduced here the notation $\Delta_{l\nu_l}(c,s)$ to signify the 
contribution of the $(c,s)$ loop to the asymmetry in the channel $l\nu_l$. 
Similarly,  $\Delta_{l\nu_l}(u,d)$ denotes the contribution of the $(u,d)$ 
loop. In deriving (17), we have used the optical theorem in the form $ImF_5
(q^2=m_H^2)=m_H\Gamma_H$, and have neglected terms of relative order $g^2$.

\subsection{Asymmetry in Quark Channels}
In complete analogy to the lepton case, the matrix elements for the 
decays $t\to bc\overline s$, $\overline t\to\overline b\overline c s$ are
\begin{eqnarray}
M=\frac{ig^2}{8}&\Bigl\{&A\overline u_b\gamma^{\mu}(1-\gamma_5)u_t
\overline u_c\gamma_{\mu}(1-\gamma_5)v_s\nonumber\\
&+&B\overline u_b(1+\gamma_5)u_t\overline u_c(1+\gamma_5)v_s
+C\overline u_b(1+\gamma_5)u_t\overline u_c(1-\gamma_5)v_s\nonumber\\
&+&D\overline u_b(1-\gamma_5)u_t\overline u_c(1+\gamma_5)v_s
+E\overline u_b(1-\gamma_5)u_t\overline u_c(1-\gamma_5)v_s\Bigr \},
\nonumber\\
\overline M=\frac{ig^2}{8}&\Bigl\{&\overline A\overline v_t\gamma^{\mu}
(1-\gamma_5)v_b\overline u_s\gamma_{\mu}(1-\gamma_5)v_c\nonumber\\
&+&\overline B\overline v_t(1-\gamma_5)v_b\overline u_s(1-\gamma_5)
v_c+\overline C\overline v_t(1-\gamma_5)v_b\overline u_s(1+\gamma_5)
v_c\nonumber\\
&+&\overline D\overline v_t(1+\gamma_5)v_b\overline u_s(1-\gamma_5)v_c
+\overline E\overline v_t(1+\gamma_5)v_b\overline u_s(1+\gamma_5)v_c
\Bigr\},
\end{eqnarray}
where
\begin{eqnarray}
A&=&G_T,\nonumber\\
B&=&\frac{m_tm_s}{m_W^2}\Bigl \{\frac{m_W^2}{q^2}(G_T+G_L)
-XY^*G_5+m_WN(Y^*F_4-XF_3)\Bigr\}, \nonumber\\
C&=&\frac{m_tm_c}{m_W^2}\Bigl \{-\frac{m_W^2}{q^2}(G_T+G_L)
-|Y|^2G_5-m_WN(Y^*F_4+YF_3)\Bigr\}, \nonumber\\
D&=&\frac{m_bm_s}{m_W^2}\Bigl \{-\frac{m_W^2}{q^2}(G_T+G_L)
-|X|^2G_5+m_WN(X^*F_4+XF_3)\Bigr\},\nonumber\\
E&=&\frac{m_bm_c}{m_W^2}\Bigl \{\frac{m_W^2}{q^2}(G_T+G_L)
-X^*YG_5+m_WN(YF_3-X^*F_4)\Bigr\}
\end{eqnarray}
and
\begin{eqnarray}
\overline A&=&A,\qquad \overline C=C,\qquad \overline D=D,\nonumber\\
\overline B&=&\frac{m_tm_s}{m_W^2}\Bigl \{\frac{m_W^2}{q^2}(G_T+G_L)
-X^*YG_5+m_WN(YF_3-X^*F_4)\Bigr\},\nonumber\\
\overline E&=&\frac{m_bm_c}{m_W^2}\Bigl \{\frac{m_W^2}{q^2}(G_T+G_L)
-XY^*G_5+m_WN(Y^*F_4-XF_3)\Bigr\}.
\end{eqnarray}
Once again , the transverse propagator term $G_T$ makes no contribution to 
the asymmetry, which is given by
\begin{eqnarray}
\Delta_{cs}&\equiv &\Gamma (\overline t\to \overline b\overline cs)-
\Gamma (t\to bc\overline s)\nonumber\\
&=&\frac{N_cg^4}{2^{11}\pi^3m_t^3}\int\limits^{(m_t-m_b)^2}_{(m_c+m_s)^2}
\frac{dq^2}{q^2}\lambda (q^2,m_t^2,m_b^2)\lambda (q^2,m_c^2,m_s^2)
\Bigl\{\nonumber\\
&&(|\overline B_0|^2-|B_0|^2+|\overline E_0|^2-|E_0|^2)(m_t^2+m_b^2-q^2)
(q^2-m_c^2-m_s^2)\nonumber\\
&&-4Re\bigl [C_0^*(\overline B_0-B_0)+D_0^*(\overline E_0-E_0)\bigr ]m_cm_s
(m_t^2+m_b^2-q^2)\nonumber\\
&&+4Re\bigl [D_0^*(\overline B_0-B_0)+C_0^*(\overline E_0-E_0)\bigr ]m_tm_b
(q^2-m_c^2-m_s^2)\Bigr\},
\end{eqnarray}
where the subscript ``$0$'' means the expressions (19) and (20) without 
the terms proportional to $G_T$. 
Expressed in terms of the functions $F_i(q^2)$, the asymmetry involves 
only the combinations given in Eq. (16), yielding as the final result
\begin{eqnarray}
\Delta_{cs}&=& \frac{N_c^2g^6}{2^{13}\pi^4m_t^3m_W^6}ImXY^*|X+Y|^2
\int\limits_{Max[(m_c+m_s)^2,(m_u+m_d)^2]}^{(m_t-m_b)^2}
\frac{dq^2}{q^6}\nonumber\\
&&\frac{\lambda (q^2,m_c^2,m_s^2)\lambda (q^2,m_u^2,m_d^2)
\lambda (q^2,m_t^2,m_b^2)}{(q^2-m_H^2)^2+m_H^2\Gamma_H^2}\cdot
f(q^2,m_t^2,m_b^2,m_c^2,m_s^2,m_u^2,m_d^2)\nonumber\\
&-& \sum_{l=e,\mu ,\tau}\Delta_{l\nu_l}(c,s),
\end{eqnarray}
where the last term follows from the relation $\Delta_{cs}(l\nu_l)=-
\Delta_{l\nu_l}(cs)$, which we have checked explicitly. 
The function $f$ is defined by
\begin{eqnarray}
f(q^2,t,b,c,s,u,d)&=& q^4(tbcd-tbsu-tcsd+tsud+bcsu-bcud)\nonumber\\
&+&q^2(t^2csd-t^2sud-tbc^2d-tbcd^2+tbs^2u+tbsu^2\nonumber\\
&&+tcsd^2-ts^2ud-b^2csu+b^2cud+bc^2ud-bcsu^2)\nonumber\\
&&+(ts-bc)(su-cd)(td-bu).
\end{eqnarray}
It has the remarkable property of being antisymmetric under any one of 
the following exchanges:
\begin{eqnarray}
(u,d)\leftrightarrow (c,s)\qquad ;\qquad (u,d)\leftrightarrow (t,b)
\qquad ;\qquad (c,s)\leftrightarrow (t,b).
\end{eqnarray}
As a consequence of this asymmetry, we immediately see that (i) the $(c,s)$ 
loop does not contribute to $\Delta_{cs }$, (ii) the analogous result 
for $\Delta_{ud}$ is obtained by interchanging $(c,s)$ and $(u,d)$ 
in Eq. (22), and (iii) the asymmetries in the various channels satisfy the 
relation
\begin{eqnarray}
\Delta_{cs}+\Delta_{ud}+\Delta_{\tau\nu_{\tau}}+\Delta_{\mu\nu_{\mu}}
+\Delta_{e\nu_e}=0,
\end{eqnarray}
implying the equality of  total width of $t$  and $\overline t$, mandated by 
$CPT$ invariance.

\section{Comments}
(i) Our results fulfil all the general constraints on partial width 
asymmetries noted by Wolfenstein \cite{wolfenstein}. In particular, the 
asymmetry in a channel $f$ associated with a loop $n$ satisfies
\begin{eqnarray}
\Delta_f(n)=-\Delta_n(f)
\end{eqnarray}
and vanishes when $n=f$.\\
(ii) A characteristic feature of $W^+-H^+$ interference is the result 
 that the asymmetry $\Delta_q(q')$ in the quark channel $q$, arising from 
a quark loop $q'$, is proportional to the function $f(q^2,m_t^2,m_b^2,
m_1^2,m_2^2,m_3^2,m_4^2)$ defined in Eq. (23), where $(m_1,m_2)$ and 
$(m_3,m_4)$ are the masses of the quark doublets contained in $q$ and 
$q'$. This implies that the specific asymmetry $\Delta_q(q')$ 
vanishes when one of the masses $(m_1,m_2)$ and one of the masses 
$(m_3,m_4)$ is zero. For a similar reason, the asymmetry in a lepton 
channel $l$ due to a lepton loop $l'$ vanishes, even for $l\ne l'$, since 
the two doublets necessarily contain two massless neutrinos.\\
(iii) The fact that the asymmetries $\Delta_{l\nu_l}$, $\Delta_{cs}$ and 
$\Delta_{ud}$ given by Eqs. (17) and (22) satisfy the $CPT$ 
condition (25) is a nontrivial test of the full $W^+-H^+$ propagator 
constructed in Eq. (8). In particular, neglect of the off-diagonal 
terms $F_3$ and $F_4$ leads to conflict with $CPT$ invariance. These 
terms have not been considered in previous work.\\
(iv) Our results for $\Delta_{\tau\nu_{\tau}}$ and $\Delta_{cs}$ 
do not coincide with those in Refs. \cite{atwood,liu}. For instance, these earlier 
papers found an asymmetry $\Delta_{\tau\nu_{\tau}}$ proportional to 
$m_{\tau}^2m_c^2$. By contrast, the leading term of our result (Eq. (17)) 
is proportional to $m_{\tau}^2m_s^2$. We have been able to trace the 
difference to the neglect of the off-diagonal part of the $W^+-H^+$ 
propagator in Refs. \cite{atwood,liu}, which inevitably leads to a 
violation of the $CPT$ condition (Eq. (25)).\\
(v) In the absence of any scalar interaction of the form ${\cal L}_H$, the 
transverse and longitudinal parts of the propagator $D_W^{\mu\nu}$ 
obtained by us agree with those in Refs. \cite{atwood,liu,nowakowski}.\\
(vi) Numerically, the partial width asymmetries resulting from 
$W^+-H^+$ interference, in the models discussed here, are exceedingly 
small. As pointed out in Ref. \cite{atwood}, larger differences between 
$t\to b\tau^+\nu_{\tau}$ and $\overline t\to\overline b\tau^-
\overline{\nu}_{\tau}$ occur if one compares the spectra of these 
reactions, not only in the variable $q^2$ but also in the complementary 
Dalitz variable $u=(p_{\tau}+p_b)^2$ \cite{atwood2,cruz}. Likewise, 
larger asymmetries are possible if one compares the $\tau^+$ and $\tau^-$ 
polarization \cite{atwood3}. Whereas the partial width asymmetry discussed 
in this paper involves only the longitudinal part of the $W$ propagator, 
these alternative effects involve the transverse part, and do not 
necessarily require absorptive phases associated with final state 
interactions.
\\
\\
\\
Acknowledgements: We wish to record our indebtedness to the papers  
of Lincoln Wolfenstein on all aspects of $CP$-symmetry, particularly Ref. 
\cite{wolfenstein}. The support of the German Ministry of Research 
and Technology is acknowledged with gratitude. One of us (T.A.) has 
been supported by the Graduiertenf\"orderungsgesetz Nordrhein-Westfalen.

\newpage
\Large
\noindent{\bf Figure Captions}
\normalsize
\begin{itemize}
\item[Fig. 1.] Diagonal and non-diagonal one-particle-irreducible 
self-energies of the $W-H$ system (Eq. (7)).
\item[Fig. 2.] Graphical representation of the ``pure'' $W$ and $H$ 
propagators, neglecting $W-H$ mixing.
\item[Fig. 3.] Graphical representation of the full $W-H$ propagator 
(Eq. (8)), in terms of the ``pure'' $W$ and $H$ propagators defined in 
Fig. 2.
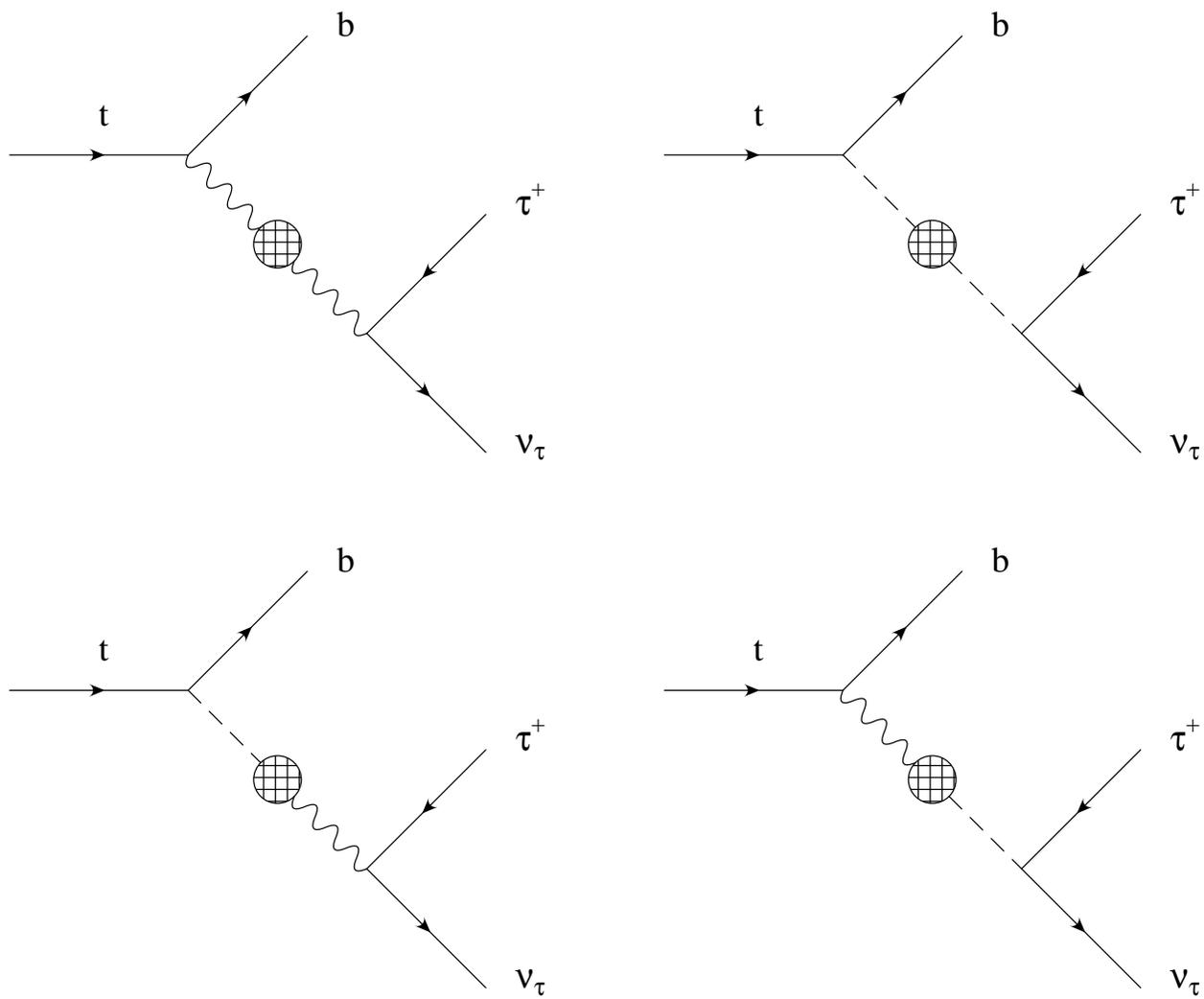
\item[Fig. 4.] Feynman diagrams contributing to the reaction 
$t\to b\tau^+\nu_{\tau}$.
\end{itemize}

\end{document}